\RequirePackage{ifpdf}
\ifpdf % We are running pdfTeX in pdf mode
\documentclass[pdftex]{sigma}
\else
\documentclass{sigma}
\fi

\begin{document}

\allowdisplaybreaks

\renewcommand{\PaperNumber}{008}

\FirstPageHeading

\renewcommand{\thefootnote}{$\star$}

\ShortArticleName{Models for Quadratic Algebras}

\ArticleName{Models for Quadratic Algebras Associated\\ with Second Order Superintegrable Systems in 2D\footnote{This
paper is a contribution to the Proceedings of the Seventh
International Conference ``Symmetry in Nonlinear Mathematical
Physics'' (June 24--30, 2007, Kyiv, Ukraine). The full collection
is available at
\href{http://www.emis.de/journals/SIGMA/symmetry2007.html}{http://www.emis.de/journals/SIGMA/symmetry2007.html}}}

% Names of the authors for the title of the paper
\Author{Ernest G.~KALNINS~$^\dag$, Willard MILLER Jr.~$^\ddag$ and Sarah POST~$^\ddag$}

\AuthorNameForHeading{E.G.\ Kalnins, W.\ Miller Jr. and S.~Post}

\Address{$^\dag$~Department of Mathematics,
 University
of Waikato, Hamilton, New Zealand}
\EmailD{\href{mailto:math0236@math.waikato.ac.nz}{math0236@math.waikato.ac.nz}}
\URLaddressD{\url{http://www.math.waikato.ac.nz}}

\Address{$^\ddag$~School of Mathematics, University of Minnesota,
 Minneapolis, Minnesota, 55455, USA}
\EmailD{\href{mailto:miller@ima.umn.edu}{miller@ima.umn.edu}, \href{mailto:postx052@math.umn.edu}{postx052@math.umn.edu}}
				% of Second Author
\URLaddressD{\url{http://www.ima.umn.edu/~miller/}}

\ArticleDates{Received October 25, 2007, in f\/inal form January
15, 2008; Published online January 18, 2008}

\Abstract{There are 13 equivalence classes of 2D second order quantum and classical superintegrable systems with nontrivial potential,
each associated with a quadratic algebra of hidden symmetries. We study the f\/inite and inf\/inite irreducible representations of the
quantum quadratic  algebras though the construction of  models in which the symmetries act on   spaces of functions of a single
complex variable via either dif\/ferential operators or dif\/ference operators. In another paper we have already carried out parts of this
analysis for the generic nondegenerate superintegrable system on the complex 2-sphere. Here we carry it out for a degenerate
superintegrable system on the 2-sphere. We point out the connection between our results and a position dependent mass Hamiltonian
studied by Quesne. We also show how to derive simple models of the classical quadratic algebras for superintegrable systems and
then obtain the quantum models from the classical models, even though the classical and quantum quadratic algebras are distinct. }


\Classification{20C99; 20C35; 22E70}

\section{Introduction}
A classical (or quantum)  $m$th order superintegrable system  is an integrable
 $n$-dimensional Hamiltonian system with potential that
admits $2n-1$ functionally independent  constants of the motion, the
 maximum possible, and such that the constants of the motion  are
 polynomial of at most order $m$ in the momenta.  Such systems are of
 special signif\/icance in mathematical physics because the trajectories
 of the classical motions can be determined by algebraic means alone,
 whereas the quantum eigenvalues for the energy and the other symmetry
 operators can also be determined by algebraic methods.
 In contrast to merely
 integrable systems, they can be solved in multiple ways. The best
 known (and historically most important) examples are the classical
 Kepler system and the quantum Coulomb (hydrogen atom) system, as well
 as the isotropic oscillator. For these examples $m=2$ and the most
 complete classif\/ication and structure results are known for the
 second order case. There is an extensive
 literature on the subject \cite{WOJ,EVA,EVAN,FMSUW,BDK,CDas,VILE,RAN,KMWP,KMJP2,KMJP3, Koenigs, CAL, WOJW,
KKMP, Zhedanov1991, Zhedanov1992a,
 Zhedanov1992b, Quesne1994,TTW,GW,HER1},  with a recent new burst of
 activity \cite{KKM20041,KKM20042,KKM20051,
 KKM20052,KKM20061,KMP2005,
 KKW,KKMW,DASK2005,HMS,SCQS,KKM2007a,KKM2007b, KKM2007c, Quesne2007,
 DASK2007,KMPost,GRAVEL,BH,FORDY,KOP2, Kress2007}.  All
 such systems have been classif\/ied for real and complex Riemannian
 spaces with $n=2$ and their associated quadratic
 algebras of symmetries computed \cite{KKMP,KKW,KKMW, KKM20042,DASK2005}. For nonconstant potentials there are
13 equivalence classes of such stems (under the St\"ackel transform between manifolds), 7 with nondegenerate (3-parameter)
potentials and
6 with degenerate (1-parameter) potentials~\cite{Kress2007,KKMW}. The constants of the motion for each system
generate a quadratic algebra
that closes at order 6 in the nondegenerate case and at order 4 in the degenerate case.

The representation theory of such algebras is of great interest
 because it is this quadratic algebra ``hidden symmetry" that accounts
 for the  degeneracies  of the energy levels of the quantum systems
 and the ability to compute all associated spectra of such systems by
 algebraic means alone.  In principle, all of these quadratic algebras can be obtained from the
quadratic algebra of  a single generic 3-parameter potential on the complex two-sphere
 by prescribed limit operations and through St\"ackel
 transforms. However, these limiting operations are not yet
 suf\/f\/iciently understood. Each
 equivalence class has special properties, and each of the  13 cases is
 worthy of study in its own right. A powerful technique for carrying
 out this study is the use of ``one variable models''. In the quantum
 case these are realizations of the quadratic algebra (on an
 energy eigenspace) in terms of dif\/ferential or dif\/ference operators
 acting on a space of functions of a single complex variable, and for
 which the energy eigenvalue is constant. Each model is adapted to the spectral
 decomposition of one of the symmetry operators, in particular, one
 that is associated with variable separation in the original quantum
 system. The possible  irreducible representations can be
 constructed on these spaces with function space inner or bilinear
 products (as appropriate) and intertwining operators to map the representation space
 to the solution space of the associated quantum system. (There have
 been several elegant treatments of the representation theory of some
 quadratic algebras, e.g.~\cite{Zhedanov1991, Zhedanov1992a,
 Zhedanov1992b,BDK,CDas, Quesne1994, Quesne2007}. However these have almost always
been restricted to f\/inite dimensional and unitary
representations and the question of determining all one variable models has not been addressed.)
In the classical case these are realizations of the
quadratic algebra (restricted to a constant energy surface) by functions of a single pair of canonical
conjugate variables.

In \cite{KMPost} we have already carried out parts of this analysis for the generic nondegenerate superintegrable
system on the complex 2-sphere.
There the potential was $V=a_1/s_1^2+a_2/s_2^2+a_3/s_3^2$ where $s_1^2+s_2^2+s_3^2=1$, and the one
variable quantum model was expressed
in terms of dif\/ference operators. It gave  exactly the algebra that describes the Wilson and Racah polynomials in
their full generality.
In this paper we treat a superintegrable case with a degenerate potential. Our example is again on the complex 2-sphere,
 but now the potential is
$V=\alpha/s_3^2$. Though this potential is a restriction of the generic potential, the degenerate case admits
a Killing vector so the quadratic algebra
structure changes dramatically. The associated quadratic algebra closes at level 4 and has a richer representation
theory than the nondegenerate case.
Now we f\/ind one variable models for an irreducible representation expressed as either  dif\/ference or
dif\/ferential operators, or sometimes both.
We show that this system can occur in unobvious ways, such as in a position dependent mass Hamiltonian recently
introduced by Quesne~\cite{Quesne2007}.

The second part of the paper concerns models of classical quadratic
algebras. Here  we inaugurate this study, in particular  its relationship to quantum models of
superintegrable systems. We f\/irst describe how these classical
models arise out of standard Hamilton--Jacobi theory. In \cite{Kress2007,KKM20061} we have shown that
for second order superintegrable systems in two
dimensions there is a 1-1 relationship between classical quadratic algebras and quantum quadratic algebras,
even though these algebras are not isomorphic.
In this sense the quantum quadratic algebra, the spectral theory for its irreducible representations and its
possible one variable models are already uniquely
determined by the classical system. We make this concrete by showing explicitly how the possible classical
models of  the classical superintegrable system with
potential $V=\alpha/s_3^2$ lead directly to the possible one variable dif\/ferential or dif\/ference operator
models for the quantum quadratic algebra. Then
we repeat this analysis for the nondegenerate potential $V=a_1/s_1^2+a_2/s_2^2+a_3/s_3^2$  where the
quantum model is essentially the Racah algebra QR(3)
and its inf\/inite dimensional extension to describe the Wilson polynomials. Our results show that the Wilson
polynomial structure is already imbedded in the
classical system with potential $V=a_1/s_1^2+a_2/s_2^2+a_3/s_3^2$, even though this potential admits
no Lie symmetries. Thus the properties of the Wilson
polynomials in their full generality could have been derived directly from classical mechanics!

This work is part of a long term project to study the structure and representation theory for quadratic algebras associated with superintegrable systems in $n$
dimensions \cite{KKM20041,KKM20042,KKM20051,KKM20052,KKM20061,KKM2007b,KKM2007c}.
The analysis for $n=3$ dimensions will be much
more challenging, but also a good indication of behavior for general $n$.

\section{The structure equations for S3}
Up to a St\"ackel transform, every 2D second order superintegrable system  with nonconstant potential is equivalent to one of 13 systems \cite{Kress2007}.
There is a representative from each equivalence class on either the complex 2-sphere or complex Euclidean space. In several papers, in particular~\cite{KKMP},
we have classif\/ied all of the constant curvature superintegrable systems, and this paper focuses on two systems contained in that list: S9 and S3.
The quadratic algebra of the generic nondegenerate system S9 was already treated in \cite{KMPost} and we will return to it again in this paper.
First we study the quadratic algebra representation theory for the degenerate potential S3.
This  one-parameter potential  2-sphere system corresponds to the potential
\[ V=\frac{\alpha}{s_3^2},
\]
where
$s_1^2+s_2^2+s_3^2=1$ is the imbedding of the   sphere in Euclidean space.
The quantum degenerate superintegrable system is
\begin{equation*} %\label{2DHam1q}
H=J_1^2+J_2^2+J_3^2+V(x,y)=H_0+V,
\end{equation*}
where      $J_3=s_1\partial_{s_2}-s_2\partial_{s_1}$  and $J_2$, $J_3$
are obtained by cyclic permutations of the indices $1,2,3$.
The basis symmetries are
\begin{gather*} L_1=J_1^2+\frac{\alpha s_2^2}{s_3^2},\!\!\!\qquad
 L_2=\frac12(J_1J_2+J_2J_1)-\frac{\alpha s_1s_2}{s_3^2},\!\!\!\qquad X=J_3,\!\!\!\qquad
 H=J_1^2+J_2^2+J_3^2+V,
 \end{gather*}
where $J_3=s_2\partial_{s_1}-s_1\partial_{s_2}$ plus cyclic
 permutations. They generate a  quadratic algebra that closes at order 4. The quadratic algebra relations are $[H,X]=[H,L_j]=0$ and
\begin{gather}\label{structure1} [L_1,X]=2L_2,\qquad
 [L_2,X]=-X^2-2L_1+H-\alpha,\\
 [L_1,L_2]=-(
 L_1X+XL_1)-\left(\frac12+2\alpha\right)X.\nonumber
 \end{gather}
The Casimir relation is \begin{gather}
 {\cal C}\equiv \frac13\left(X^2L_1+XL_1X+L_1X^2\right)+L_1^2+L_2^2-HL_1+\left(\alpha+\frac{11}{12}\right)X^2-\frac16
 H\nonumber \\
\phantom{{\cal C}\equiv}{} +\left(\alpha-\frac23\right)L_1-\frac{5\alpha}{6}=0.\label{structure2} \end{gather}

We know that the quantum Schr\"odinger equation separates in
spherical coordinates, and that corresponding to a f\/ixed energy
eigenvalue $H$, the eigenvalues of $X$ take the linear  form
\begin{equation*}%\label{evs1}
\lambda_n={\cal A}n+{\cal B},
\end{equation*}
where $n$ is an integer, so we will look for irreducible
representations of the quadratic algebra such that the representation
space has a basis of eigenvectors $f_n$ with corresponding eigenvalues
$\lambda_n$.  (Indeed, from the analysis of  \cite{Zhedanov1991} or of \cite{CDas} the structure equations imply
that the spectrum of~$X$ must be of this form.)  We will use the
abstract structure equations to list the corresponding representations
and compute the action of $L_1$ and $L_2$ on an $X$ basis. Thus,  we
assume that there is a basis $\{f_n\}$,  for the representation space such that
\begin{equation*}%\label{basis1}
Xf_n=\lambda_nf_n,\qquad  L_1f_n=\sum_j C(j,n)f_j,\qquad  L_2f_n=\sum_j D(j,n)f_j.\end{equation*}
Here, $\cal A$, $\cal B$ are  not yet f\/ixed. We do not impose any inner product
space structure.

 From these assumptions we can compute the action of $L_1$ and $L_2$  on the basis. Indeed,
\begin{gather}\label{calc1} [L_1,L_2]f_n=\sum_{j,k} \left(C(j,k)D(k,n)-D(j,k)C(k,n)\right)f_j,\\
\label{calc2}[L_1,X]f_n=\sum_j (\lambda_n-\lambda_j)C(j,n)f_j,\qquad [L_2,X]f_n=\sum_j (\lambda_n-\lambda_j)D(j,n)f_j
.\end{gather}
On the other hand, from the  equations \eqref{structure1}   we
have
\begin{gather}\label{structure3a} [L_1,X]f_n
=2\sum_jD(j,n)f_j,
\\
\label{structure3b}[L_2,X]f_n
=-2\sum_jC(j,n)f_j+(-\lambda_n^2+H-\alpha)f_n,\\
\label{structure3c}
[L_1,L_2]f_n=-\sum_j(\lambda_n+\lambda_j)C(j,n)f_j-\left(\frac12+2\alpha\right)\lambda_nf_n.
\end{gather}

Now we equate equations \eqref{calc2} with \eqref{structure3a} or \eqref{structure3b}. For $j=n$, equating
 coef\/f\/icients of $f_n$ in the resulting identities yields the
 conditions
\begin{equation*}%\label{CDdiag}
D(n,n)=0, \qquad C(n,n)=\frac{-\lambda_n^2+H-\alpha}{2}.
\end{equation*}
Similarly, equating coef\/f\/icients of $f_j$ in the case $j\ne n$ yields
\[{\cal A}(n-j)D(j,n)=-2C(j,n),\qquad {\cal A}(n-j)C(j,n)=2D(j,n),
\]
or
\[ \left({\cal A}^2(n-j)^2+4\right) C(j,n)=0,\qquad j\ne n.
\]
Thus, either $C(j,n$ and $D(j,n)$ vanish or ${\cal A}^2(n-j)^2=-4$. We can
 scale $\cal A$ such that the smallest nonzero jump is for $j=n\pm
1$, in which case ${\cal A}=\pm 2i$. By replacing $n$ by $-n$ if necessary,
we can assume ${\cal A}=2i$. (We also set ${\cal B}=i\mu$.) Thus the only possible nonzero values of
$C(j,n)$, $D(j,n)$ are for $j=n,n\pm 1$ and there are the relations
\begin{equation*}%\label{CDrelations}
 D(n+1,n)=-iC(n+1,n),\qquad D(n-1,n)=iC(n-1,n).
\end{equation*}

Comparing \eqref{calc1} and \eqref{structure3c} and equating coef\/f\/icients of~$f_{n\pm 2}$, $f_{n\pm 1}$, respectively,  on both sides of the resulting
  identities, we do not obtain new conditions. However, equating
  coef\/f\/icients of~$f_n$ results in the condition
\begin{equation*}%\label{calc3}
F_{n+1}-F_n= \frac12 (2n+\mu)\left(4n^2+4\mu n+\mu^2+H+\alpha+\frac12\right),
\end{equation*}
where $F_n=C(n,n-1)C(n-1,n)$.
 The  general solution of this dif\/ference equation is
\begin{gather*}%\label{cnn+1ident}
F_n=n^4+(2\mu-2)n^3+\left(\frac32\mu^2-3\mu+\frac{H}{2}+\frac{\alpha}{2}+\frac53\right)n^2\\
\phantom{F_n=}{} +\left(\frac{\mu^3}{2}-\frac{3\mu^2}{2}+\left[\frac{H}{2}+\frac{\alpha}{2}+\frac54\right]\mu-\frac{H}{2}-\frac{\alpha}{2}-\frac14\right)n
+\kappa,\nonumber\end{gather*}
where $\kappa$ is an arbitrary constant.

To determine $\kappa$ we substitute these results into the Casimir
equation \eqref{structure2} and set equal to zero the  coef\/f\/icients of $f_j$ in
the expression ${\cal C}f_n=0$. For $j\ne n$ we get nothing new. However,
$j=n$ we f\/ind
\begin{gather} \kappa=
\frac{1}{16}\mu^4-\frac14\mu^3+\frac{H+\alpha+5/2}{8}\mu^2-\frac{1/2+H+\alpha}{4}\mu
\nonumber\\
\phantom{\kappa=}{} +\frac18H+\frac{\alpha}{8}+\frac{H^2}{16}-\frac{H\alpha}{8}+\frac{\alpha^2}{16}.\label{kappa}\end{gather}
Thus, $F_n=C(n,n-1)C(n-1,n)$ is an explicit 4th order polynomial in
$n$. By factoring this polynomial in various ways, and re-normalizing
the basis vectors $f_n$ appropriately via $f_n\to c(n)f_n$,  we can achieve a realization
of the action of $L_1$ and $L_2$ such that
\begin{gather} L_1f_n=C(n+1,n)f_{n+1}+ C(n,n)f_{n} + C(n-1,n)f_{n-1},\nonumber\\
L_2f_n=D(n+1,n)f_{n+1}+ D(n,n)f_{n} + D(n-1,n)f_{n-1}\label{Ldef}
\end{gather}
and all of the coef\/f\/icients are polynomials in $n$. The 4 roots of $F_n$
are
\[ \frac{1-\mu}{2}\pm\frac{1}{4}\sqrt{2-4(H+\alpha)\pm
  2\sqrt{1-4(H+\alpha)+16H\alpha}},\]
so a convenient factorization is
\begin{gather*}%\label{fact1}
C(n,n-1)=\left(n+\frac{\mu- 1}{2}\right)^2-\frac{1}{8}\left(1-2(H+\alpha)+
  \sqrt{1-4(H+\alpha)+16H\alpha}\right),
\\
 C(n-1,n)=\left(n+\frac{\mu-1}{2}\right)^2-\frac{1}{8}\left(1-2(H+\alpha)-
  \sqrt{1-4(H+\alpha)+16H\alpha}\right).
\nonumber\end{gather*}
From these expressions and from
\begin{gather*}%\label{CDdiag1}
D(n,n)=0, \qquad C(n,n)=\frac{(2n+\mu)^2+H-\alpha}{2},\qquad
D(n\pm 1,n)=\mp iC(n\pm 1,n)
\end{gather*}
we see that we can  f\/ind $C$, $D$ coef\/f\/icients in which the the dependence on $n$ is always as a~polynomial.

There are raising and lowering operators
\begin{equation*}%\label{raising}
A^\dagger=L_1+iL_2+\frac12(X^2-H+\alpha),\qquad
 A=L_1-iL_2+\frac12(X^2-H+\alpha).\end{equation*}
Indeed,
\[A^\dagger f_n=2C(n+1,n)f_{n+1},\qquad Af_n=2C(n-1,n)f_{n-1},\]
and $[A,A^\dagger]f_n=2(F_{n+1}-F_n)f_n$, so $[A,A^\dagger]$ is a third order polynomial in $X$.

To get a
one-variable model of the quadratic algebra in terms of second order
dif\/ferential operators, we can simply make the choices $f_n(t)=t^n$,
$X=i(2t\frac{d}{dt}+\mu)$ and def\/ine $L_1$  from expressions
\eqref{Ldef} via the prescription
\begin{equation}\label{Ldef1} L_1f_n(t)=\left(tC\left(t\frac{d}{dt}+1,t\frac{d}{dt}\right)+ C\left(t\frac{d}{dt},t\frac{d}{dt}\right) +
t^{-1}C\left(t\frac{d}{dt}-1,t\frac{d}{dt}\right)\right)f_{n}(t),\end{equation}
with a similar procedure for $L_2$.

In general the irreducible  representations that we have def\/ined are
inf\/inite dimensional and the basis vectors $f_n$ occur for all
positive and negative
integers $n$. We can obtain representations bounded below, and with
lowest weight $\mu$ for $-iX$ and corresponding lowest weight vector $f_0$,
simply by requiring $F_0=0$, which amounts to setting $\kappa=0$. For
convenience we set $\alpha=1/4-a^2$. Then we have
\begin{equation}\label{bddbelow1} F_n=C(n,n-1)C(n-1,n)= n(n+\mu-1)(n+\mu-1+a)
(n- a).\end{equation}
%$$\ S=\sqrt{-\mu^2+2\mu-2\alpha-2H}.
%$$
Since $\kappa=0$, \eqref{kappa}, $H$ must be a solution of this  quadratic equation:
\begin{equation}\label{kappa0}
H=-(\mu-1+a)^2+\frac14.
\end{equation}
%\frac{1}{16}\mu^4-\frac14\mu^3+\frac{H+\alpha+5/2}{8}\mu^2-\frac{1/2+H+\alpha}{4}\mu
%\ee
%$$+\frac18H+\frac{\alpha}{8}+\frac{H^2}{16}-\frac{H\alpha}{8}+\frac{\alpha^2}{16}=0.$$
A convenient choice is
\begin{gather*}%\label{fact2}
C(n-1,n)=n(n+\mu-1+a),\qquad
C(n+1,n)=(n+\mu)(n+1-a),\\
 C(n,n)=2n^2+2n\mu-\mu a+a+\mu-\frac12.\nonumber\end{gather*}
If $\mu$ is not a negative integer then this bounded below
representation is inf\/inite dimensional. However, if there is a highest
weight vector $f_m$ then we must have $F_{m+1}=0$, or $\mu=-m$,
$m=0,1,\dots$.
Thus the f\/inite dimensional representations are indexed by the
nonnegative integer $m$ and the eigenvalues of $-iX$ are $m,
m-2,\dots, -m$. The dimension of the representation space is
$m+1$.

At this point it is worth pointing out that {\it all} of the f\/inite dimensional, inf\/inite
dimensional  bounded below, and general inf\/inite dimensional irreducible
representations and models  of the quadratic algebra associated with the superintegrable
system S3 are of direct interest and applicability. A similar argument was made
in \cite{KMPost} where we gave examples of various ana\-ly\-tic function expansions
and distinct unitary structures associated with one superintegrable system. The original
S3 quantum system is given in terms of complex variables with no specif\/ic inner product
structure imposed. One could use the representation
theory results to describe eigenfunction expansions simply in terms of analytic functions.
If one wants an inner product structure or bilinear product structure, it is merely necessary to impose the
structure on a~{\it single} eigenspace of $H$, and there are a variety of ways to do this. For example,
one could restrict the complex system to the real sphere and impose
the standard inner product for that case. Alternatively, one could restrict to the real hyperboloid
of one sheet, or the real hyperboloid
of two sheets. In all cases the models of the irreducible representations are relevant, though not
necessarily to one special case, such as the real sphere with the standard
inner product. While we have no direct proof that all models of irreducible representations of quadratic algebras obtained
in this way lead to representations for some version of the original superintegrable system,
we have no counterexamples. For a deeper analysis we need to construct intertwining operators that relate
basis functions for the model with  eigenfunctions of the quantum Hamiltonian.

\section[Differential operator models]{Dif\/ferential operator models}

A convenient realization  of the f\/inite dimensional representations by  dif\/ferential operators in one complex variable is
\begin{gather}
 L_1=\left(t^3+2t^2+t\right)\frac{d^2}{dt^2}+
\left((2-a-m)t^2+2(1-m)t+a-m\right)\frac{d}{dt}
\nonumber\\
\phantom{L_1=}{}+m(a-1)t+a(m+1)-m-\frac12
 ,\qquad X=i(2t\frac{d}{dt}-m),\nonumber\\
\ L_2= i\left(-t^3+t\right)\frac{d^2}{dt^2}+
i\left((a+m-2)t^2+a-m\right)\frac{d}{dt}-im(a-1)t.\label{model0}\end{gather}
This model is also correct for inf\/inite dimensional bounded-below
representations, except that now the lowest weight is $\mu=-m$ where
$m\ne 0,1,2,\dots$ is a complex number. The raising and lowering
operators for the model are
\[ A^\dagger=  2t^3\frac{d^2}{dt^2}+
2(2-a-m)t^2\frac{d}{dt}+2m(a-1)t,\qquad A= 2t\frac{d^2}{dt^2}+
2(a-m)\frac{d}{dt}.
\]
In the f\/inite dimensional case, for example, the eigenvalues of $L_1$
are
\begin{equation}\label{L1ev}\chi_n=a^2-\frac14-\left(n-a+\frac12\right)^2,\qquad
n=0,1,\dots,m,
\end{equation}
and the corresponding unnormalized eigenfunctions are
\[
(1+t)^n{}_2F_1\left(\begin{array}{cc} n-a&n-m\\ a-m\end{array}; -t\right).
\]

Now, motivated by the quantum mechanical system on the real 2-sphere,  we impose a Hilbert space structure on the
irreducible representations such that $L_1$ and $L_2$ are self-adjoint
and $X$ is skew adjoint:
\[
\langle L_j f_n,f_{n'}\rangle = \langle f_n,L_j f_{n'}\rangle,\quad j=1,2;\qquad
\langle Xf_n,f_{n'}\rangle =-\langle f_n,Xf_{n'}\rangle.
\]

Writing $\phi_n=k_nf_n$ where $\phi_n$ has norm $1$, we have the recursion
relation
\[
k_n^2=\frac{(n-1+\mu)(n-a)}{n(n-1+\mu+a)}k_{n-1}^2.
\]
For inf\/inite dimensional bounded below representations
$ k_n^2$  must be positive for all integers $n\ge 0$, and we normalize $k_0=1$.
Thus
\[
k_n^2=\frac{(\mu)_n(1-a)_n}{n!(a+\mu)_n}.
\]

For f\/inite dimensional representations we have $\mu=-m$.
Normalizing $k_0=1$, (possible for $a<1$ or for $a>m$), we f\/ind that an orthonormal basis in the one variable model is given by
$\phi_n(t)=k_nf_n(t)=k_nt^n$, $n=0,1,\dots,m$ where
\[k_n=\sqrt{\frac{(-m)_n(1-a)_n}{n!(-m
    +a)_n}}=\sqrt{\frac{m!(1-a)_n(1-a)_{m-n}}{(1-a)_m n!(m-n)!}}
.\]
Note the ref\/lection symmetry $ ||f_n||=||f_{m-n}||$.

 To derive  a realization of the Hilbert space for the dif\/ferential operator models of the f\/inite dimensional and
 inf\/inite dimensional bounded below unitary representations  in terms of a function
 space inner product
\[
\langle p,q\rangle =K\iint p(t)\overline{q(t)}\rho(t\overline {t})\, dt\,
 d\overline {t},
 \]
where $p$, $q$ are polynomials and $K$ is a normalization constant, we
 use the formal self- and skew-adjoint requirements and obtain a
 dif\/ferential equation for the weight function:
\[
(-\zeta^2+\zeta)\frac{d^2\rho(\zeta)}{d\zeta^2}+(-\mu-a+1+(-1+\mu-a)\zeta)\frac{d\rho(\zeta)}{d\zeta}+(-2+\mu-2a+a\mu)\rho(\zeta)=0,
\]
where $\zeta=t\overline{t}$. The solution that vanishes at $\zeta=1$
for $ a <1/2$  and is integrable at $\zeta=0$ for $a+\mu>-1$ is
\[\rho_1(\zeta)=(1-\zeta)^{1-2a}{}_2F_1\left(\begin{array}{ll}-\frac{\mu+3a-Q}{2}+1,&-\frac{\mu+3a+Q}{2}+1\\
2-2a\end{array};1-\zeta\right),\]
where $Q=\sqrt{a^2+(2\mu-8)a+\mu^2+4\mu-8}$. (Note that the integral
is an even function of $Q$.) At  $\zeta=0$ this
function has a branch point with behavior $\zeta^{a+\mu}$. We write
$t=re^{i\theta},\ {\overline t}=re^{-i\theta}$,  $\zeta=r^2$ and
choose our contours of integration for the inner product as the unit
circle $|e^{i\theta}|=1$, i.e., $0\le \theta\le 2\pi$ and, in the
complex $\zeta$-plane, a contour that starts at $\zeta=1$ and travels
just above the real $\zeta$-axis to circle $\zeta=0$ once in the
counterclockwise direction and returns to $\zeta=1$ just below the real
$\zeta$-axis. We require that $\langle 1,1\rangle =1$. By choosing a regime where
$a+\mu>-1$ we can shrink the $\zeta$-contour about $\zeta=0$ so that
the norm takes the form
\begin{gather*}
\langle 1,1\rangle =-4\pi K_1e^{i\pi (a+\mu)}\sin[\pi(a+\mu)]\int_0^1\rho_1(\zeta)d\zeta
\\
\phantom{\langle 1,1\rangle}{} =
\frac{-\pi K_1}{a-1}e^{i\pi (a+\mu)}\sin[\pi(a+\mu)] {}_2F_1\left(\begin{array}{ll}\frac{-\mu-3a+Q}{2}+1&\frac{-\mu-3a-Q}{2}+1\\
3-2a\end{array};1\right)\\
\phantom{\langle 1,1\rangle}{}
=4\pi K_1e^{i\pi (a+\mu)}\sin[\pi(a+\mu)]\frac{\Gamma(2-2a)\Gamma(a+\mu+1)}{\Gamma(2-\frac{a-\mu+Q}{2})\Gamma(2-\frac{a-\mu-Q}{2})},
\end{gather*}
where we have integrated term-by-term and then made use of Gauss'
Theorem for the summation of ${}_2F_1(1)$. This gives us the value for
$K_1$ such that $\langle 1,1\rangle =1$. Now, the result extends for the original
contour by analytic
continuation. This def\/ines a pre-Hilbert space inner product that then
can be extended to obtain a true Hilbert space.

The contour integral for the inner product obtained in the previous
paragraph requires \mbox{${\rm Re}\, a<1$} for convergence, and this doesn't
  hold for some of the unitary irreducible representations def\/ined
  above. Accordingly, we consider a second solution of the weight
  function equation. The solution that vanishes at $\zeta=0$
for $ a+\mu>0$  and is integrable at $\zeta=1$ for $a<1$ is
\[\rho_2(\zeta)=\zeta^{\mu+a}{}_2F_1\left(\begin{array}{ll}-\frac{\mu+3a-Q}{2},&-\frac{\mu+3a+Q}{2}\\
\mu+a+1\end{array};\zeta\right).\] At $\zeta=1$ this
function has a branch point with behavior $(1-\zeta)^{1-2a}$. We write
$t=re^{i\theta}$, ${\overline t}=re^{-i\theta}$,  $\zeta=r^2$ and
choose our contours of integration for the inner product as the unit
circle $|e^{i\theta}|=1$ and, in the
complex $\zeta$-plane, a contour that starts at $\zeta=0$ and travels
just below the real $\zeta$-axis to circle $\zeta=1$ once in the
counterclockwise direction and returns to $\zeta=0$ just above the real
$\zeta$-axis. This integral converges for ${\rm Re}\, (a+\mu)>-1$. We require that $\langle 1,1\rangle =1$. By choosing a~regime where
$a<1$ we can shrink the $\zeta$-contour about $\zeta=1$ so that
the norm takes the form
\begin{gather*}
\langle 1,1\rangle =4\pi K_2e^{i\pi (2a-1)}\sin[\pi(2a-1)]\int_0^1\rho_2(\zeta)d\zeta
\\
\phantom{\langle 1,1\rangle}{} =
\frac{\pi K_2}{\mu+a+1}e^{i\pi (2a-1)}\sin[\pi(2a-1)] {}_2F_1\left(\begin{array}{ll}\frac{\mu+3a+Q}{2}&\frac{\mu+3a-Q}{2}\\
a+\mu+2\end{array};1\right)\\
\phantom{\langle 1,1\rangle}{}
=-4\pi K_2e^{i\pi (2a-1)}\sin[\pi(2a-1)]\frac{\Gamma(2-2a)\Gamma(a+\mu+1)}
{\Gamma(2-\frac{a-\mu+Q}{2})\Gamma(2-\frac{a-\mu-Q}{2})}.
\end{gather*}
 This gives us the value for
$K$ such that $\langle 1,1\rangle =1$, and  the result extends by analytic
continuation to all values of $a$, $\mu$ for which the original contour
integral converges.

Thus we have an explicit  pre-Hilbert function space inner product for each of our
dif\/ferential operator models. In the f\/inite dimensional case we have the
 reproducing kernel function
\[
\delta(t,{\overline
   s})=\sum_{n=0}^m\phi_n(t){\overline\phi_n(s)}={}_2F_1\left(\begin{array}{cc}-m&1-a\\ -m+a\end{array} ; t\overline{s}\right).
\]
 In the inf\/inite dimensional bounded below case we have the
 reproducing kernel function
\[
\delta(t,{\overline
   s})=\sum_{n=0}^\infty\phi_n(t){\overline\phi_n(s)}={}_2F_1\left(\begin{array}{cc}\mu&1-a\\ \mu+a\end{array} ; t\overline{s}\right)
\]
which converges as an analytic function and in the Hilbert space norm
   for $|s|<1$. Here,
\[||\delta(s,{\overline s})||={}_2F_1\left(\begin{array}{cc}\mu&1-a\\ \mu+a\end{array} ; |s|^2\right).\]
In each case $\langle f(t),\delta(t,{\overline
   s})\rangle =f(s)$ for $f$ in the Hilbert space.

\section[Difference operator models]{Dif\/ference operator models}

There are also  dif\/ference operator models for the representations of
the S3 quadratic algebra. We f\/irst give the details for the f\/inite
dimensional representations indexed by the nonnegative integer
$m$. Here the operator $L_1$ is diagonalized:
\begin{gather*}L_1=-\lambda(t)+a-\frac12,\qquad \lambda(t)=t(t-2a+1),\nonumber\\
%\label{difference1}
-iX=\frac{(t-2a+1)(t-m)}{2t-2a+1}T^1-\frac{t(t+m-2a+1)}{2t-2a+1}T^{-1},
\\
L_2=\frac{(t-a+1)(t-2a+1)(t-m)}{2t-2a+1}T^1+\frac{t(t-a)(t+m-2a+1)}{2t-2a+1}T^{-1},\nonumber\end{gather*}
where $T^k$ is the dif\/ference operator $T^kf(t)=f(t+k)$.  The basis
functions are $f_n(t)=(-1)^np_n(\lambda)$ where
\[
p_n(\lambda(t))={}_3F_2\left(\begin{array}{lll}-n&-t&t-2a+1\\-m&1-a\end{array};1\right).\]
Here $f_n$ is a polynomial of order $n$ in the variable $\lambda(t)$,
a special case of the family of dual Hahn polynomials~\cite[p.~346]{AAR}. These
polynomials are orthogonal with respect to a measure with support at
the values $t=0,1,\dots,m$, in agreement with equation \eqref{L1ev} for
the eigenvalues of $L_1$. Indeed, we have (for $a<1$)
\begin{gather*}
\sum_{t=0}^m\frac{(1-2a)_t(3/2-a)_t(-m-1)_t(-1)^t}{(1/2-a)_t(2+m-2a)_tt!}p_n(\lambda(t))p_{n'}(\lambda(t))
=\frac{(2-2a)_m(a-m)_nn!}{(1-a)_m(1-a)_n(-m)_n}\delta_{nn'}.
\end{gather*}

For the inf\/inite dimensional, bounded below,  case we have
\begin{gather}L_1=t^2+a^2-\frac14,\nonumber\\
-iX=\frac{(1/2-a-it)(\mu+a-1/2-it)}{2t}T^i-\frac{(1/2-a+it)(\mu+a-1/2+it))}{2t}T^{-i},\nonumber\\
L_2=-i\frac{(1-2it)(1/2-a-it)(\mu+a-1/2-it)}{4t}T^i\nonumber\\
\phantom{L_2=}{} -i\frac{(1+2it)(1/2-a+it)(\mu+a-1/2+it))}{4t}T^{-i}.\label{difference2}
\end{gather}
  The basis
functions are $f_n(t)=(-1)^ns_n(t^2)$ where
\begin{equation*}%\label{contdualHahn}
s_n(t^2)={}_3F_2\left(\begin{array}{lll}-n&\frac12-a+it&\frac12-a-it\\\mu&1-a\end{array};1\right).\end{equation*}
Here $f_n$ is a polynomial of order $n$ in the variable $t^2$,
a special case of the family of continuous dual Hahn polynomials~\cite[p.~331]{AAR}.
The orthogonality and normalization are  given by
\begin{gather*}%\label{contdualhahnorthog}
\frac{1}{2\pi}\int_0^\infty
  \left|\frac{\Gamma(1/2-a+it)\Gamma(\mu+a-1/2+it)\Gamma(1/2+it)}{\Gamma(2it)}\right|^2s_n(t^2)s_{n'}(t^2)\,
  dt \\
\qquad{}=\frac{\Gamma(n+\mu)\Gamma(n+1-a)\Gamma(n+\mu+a)n!}{(\mu)_n^2|(1-a)_n|^2}\delta_{nn'},\nonumber\end{gather*}
where  $\mu>1/2-a>0$.

In summary, we have found the following possibilities for bounded
below irreducible representations such that $L_1$, $L_2$ are self-adjoint
and $X$ is skew adjoint, together with associated one variable models. (Here, $n_0$ is a positive integer.)
\[\begin{array}{|c|c|c|} {\rm representation}\ &{\rm parameter\ range}&{\rm model}\\
\hline
\tsep{1ex}
{\rm f\/inite \ dimensional}&\mu=-m,\ m=0,1,2,\dots
& {\rm dif\/ferential\ operators}\\
&{\rm either}\ a<1\ {\rm or}\ a+\mu>0&{\rm dif\/ference\ operators}\bsep{1ex} \\
{\rm inf.\ dim.\ bdd.\ below}&\mu>0& {\rm dif\/ferential\ operators}\\ &a<1\ {\rm and}\ a+\mu>0&{\rm dif\/ference\ operators}\bsep{1ex}\\
 {\rm inf,\ dim.\ bdd.\ below}
&0>\mu=-n_0+t,\ t\in (0,1)& {\rm dif\/ferential\ operators}\\ & a=n_0+s,\ s\in (0,1)& \bsep{1ex} \\ {\rm inf.\ dim.\ bdd.\ below}
&0>\mu=-n_0+t,\ t\in (0,1)& {\rm dif\/ferential\ operators}\\ & -t<a<1-t& \end{array}
\]
%In all but the last case, $a,\mu$ are real.
%\\  {\rm inf,\ dim.\ bdd.\ below}   &\mu>0& {\rm dif\/ference\ operators }   \\ &a=(1-\mu)/2+i\gamma,\ \gamma\in R

\section[Quesne's position dependent mass (PDM)  system
in a two-dimensional semi-infinite layer]{Quesne's position dependent mass (PDM)  system\\
in a two-dimensional semi-inf\/inite layer}

In \cite{Quesne2007} Quesne considered a superintegrable
exactly solvable position dependent mass (PDM)  system
in a two-dimensional semi-inf\/inite layer. Her
system is equivalent via a gauge transformation to a standard quantum
mechanical problem on the real 2-sphere with potential of the form
S3. Indeed,
in Quesne's paper we are given the Hamiltonian
\[-H_Q=\cosh^2qx(\partial_x^2+\partial_y^2)+2q \cosh qx \sinh qx \partial_x +q^2 \cosh^2 qx-\frac{q^2k(k-1)}{\sinh^2qx}.\]
We adopt coordinates on the unit sphere as
\[s_1=\frac{\sin qy}{\cosh qx},\qquad
s_2=\frac{\cos qy}{\cosh qx},\qquad
s_3=\tanh qx,
\]
where  $s_1^2+s_2^2+s_3^2=1$ and the metric is $ds^2=q^2(dx^2+dy^2)/\cosh^2qx$.
The Laplacian becomes \[
\Delta_S=\frac{\cosh^2 qx}{q^2} (\partial_x^2+\partial_y^2).
\] In these coordinates, the degenerate superintegrable system S3  becomes
\[ H_{S}=\frac{\cosh^2 qx}{q^2} (\partial_x^2+\partial_y^2)+\frac{\frac{1}{4}-\alpha^2}{\tanh^2 qx}.\]
 By a gauge transform  $H_O=(\cosh qx)^{-1}H_{S} \cosh{qx}$, we get
\[H_O=\frac{1}{q^2}\left\{\cosh^2 qx(\partial_x^2+\partial_y^2)+2q\cos qx \sinh qx \partial_x+q^2\cosh^2qx+q^2\frac{\frac{1}{4}-\alpha^2}{\sinh^2qx} \right\}+\frac{1}{4}-\alpha^2.\]
Thus we have $H_Q=-q^2H_0+q^2({1}/{4}-\alpha^2)=-q^2({\cosh qx})^{-1}H_{S} \cosh{qx}+q^2({1}/{4}-\alpha^2)$, with ${1}/{4}-\alpha^2=-k(k-1)$ which has solutions $k=a+{1}/{2}$ or $k=-a+{1}/{2}$. Since $k$ is assumed positive and $a$ is required to be less than 1, we take $a<0$.

Suppose we f\/ind an eigenvector for $H_{S}$ with eigenvalue $\lambda_{S}$, call it $v_{\lambda_{S}}$. Then $\lambda_Q$ will be the eigenvalue of $v_{\lambda_Q}$ for $H_Q$. We have the transformations $v_{\lambda_Q}={v_{\lambda{S}}}/{\cosh qx}$ and $\lambda_Q=-q^2\lambda_{S}+q^2({1}/{4}-\alpha^2)=-q^2\lambda_{S}-q^2k(k-1)$.
Checking the two eigenvalues, we have $\lambda_{S}=-(\mu-1+a)^2+{1}/{4}$ and $\lambda_{Q}=q^2(N+2)(N+2k+1)$. We note that these two values coincide when  $-\mu=m=N+1$ with $m$ an integer.

Using the above calculations and the eigenfunctions given in the paper, we can obtain eigenfunctions for the S3 case as
\[v_{\lambda_{S}}=N^{(k)}_{n,\ell}(\tanh qx)^{-a+\frac{1}{2}}(\cosh qx)^{-\ell-1}P_n^{-a,\ell+1}(-\tanh^2 qx)\chi_\ell(y),\]
or in coordinates on the sphere
\[v_{\lambda_{S}}=N^{(k)}_{n,\ell}(s_3)^{-a+\frac{1}{2}}(s_1^2+s_2^2)^{\frac{\ell+1}{2}}P_n^{-a,\ell+1}(-s_3^2)\chi_\ell(y),\]
where $m=2n+\ell+1$, and
$\chi_\ell(y)=\sin[(\ell+1)qy]$ or $\cos[(\ell+1)qy]$. We can rewrite these by noting ${1}/{\cosh^2qx}=1-\tanh^2qx$ so that we can write $\sin qy={s_1}/{\sqrt{1-s_3^2}}$ and $\cos qy={s_2}/{\sqrt{1-s_3^2}}$,  then we obtain
\[\chi_\ell(y)=a_n T_{\ell+1}\left(\frac{s_2}{\sqrt{s_1^2+s_2^2}}\right)+b_n\frac{s_1}{\sqrt{s_1^2+s_2^2}} U_{\ell}
\left(\frac{s_2}{\sqrt{s_1^2+s_2^2}}\right),
\]
where $T_\ell$ and $U_\ell$ are the Chebyshev polynomials of the f\/irst and second kind, respectively.

Quesne found the S3 quadratic algebra (which closes at order 4)  but
did not use it for spectral
analysis purposes because her problem involved a boundary condition
that broke the full quadratic algebra symmetry. Instead she
considered her system as a special case of S9 and used the more
complicated S9 symmetry algebra that closes at order 6 to f\/ind the
f\/inite dimensional representations. (Note that although the 1-parameter S3 potential is a limit of the 3-parameter S9 potential as two of the parameters go to $0$,
a discontinuity occurs in the structure of the quadratic algebra. A f\/irst order symmetry appears and the number of second order symmetries jumps from~3 to~4.)
Quesne's point of
view has merit, but it complicates the spectral analysis of the
problem, since the only one-variable model is in terms of dif\/ference
operators and Racah polynomials. From  our vantage point of one
one variable dif\/ferential operator analysis for the model, Quesne's boundary conditions amount to decomposing an
irreducible subspace corresponding to an $m$-dimensional
representation into a direct sum of even and odd parity
subspaces~$V^+$, $V^-$. (Indeed her boundary conditions require choice of $\chi_\ell(y)$ in the cosine form for~$\ell$ even and in the sine form for $\ell$ odd.) Let $P$ be the operator $Pf(t)=t^{m}f(1/t)$. Since $P^2=I$
and   $ ||f_n||=||f_{m-n}||$, it is clear that $P$ is unitary. We
def\/ine unit vectors
$\Phi^+_\ell=2^{-1/2}(\phi_\ell+(-1)^m\phi_{m-\ell})$ and
$\Phi_\ell^-=2^{-1/2}(\phi_\ell-(-1)^m\phi_{m-\ell})$
for $\ell=0,1,\dots,[m/2]$. Then for $m=2k$ the vectors
$\Phi^+_\ell$, $\ell=0,\dots,k$ form an orthonormal basis for $V_m^+$ and the
vectors $\Phi^-_\ell$, $\ell=0,\dots,k-1$ form an {\it on} basis for~$V_m^-$.
For $m=2k-1$, the vectors $\Phi^+_\ell$, $\ell=0,\dots,k-1$ form an ON basis for $V_m^+$ and the
vectors $\Phi^-_\ell$, $\ell=0,\dots,k-1$ form an orthonormal basis for
$V_m^-$. These basis vectors are very easily expressible  in terms of
the one variable dif\/ferential operator model, where they are sums of
two monomials. The basis used by Quesne corresponds to the~$V^-$ subspaces.Thus our models can be used to carry
out the spectral analysis for this PDM system, and they yield a~simplif\/ication.

\subsection{Classical models for S3}
Now we describe how the methods of classical mechanics lead directly to the quantum models. The classical  system S3 on the 2-sphere is determined
by the Hamiltonian
\begin{equation*}%\label{Hamiltonian}
{\cal H}={\cal J}_1^2+{\cal J}_2^2+{\cal J}_3^2+\frac{\alpha(s_1^2+s_2^2+s_3^2)}{s_3^2},
\end{equation*}
where
${\cal J}_1=s_2p_{3}-s_3p_{2}$ and ${\cal J}_2$, ${\cal J}_3$ are cyclic permutations of this expression. For computational convenience we have imbedded the 2-sphere in Euclidean 3-space. Thus we use the Poisson bracket
\[\{{\cal F}, {\cal G}\}=\sum_{i=1}^3(-\partial_{s_i}{\cal F}\partial_{p_i}{\cal G}+\partial_{p_i}{\cal F}\partial_{s_i}{\cal G})\]
for our computations, but at the end we restrict to the sphere $s_1^2+s_2^2+s_3^2=1$. The classical basis for the constants of the motion is
\begin{equation*}%\label{constants}
{\cal L}_1={\cal J}_1^2+\alpha\frac{s_2^2}{s_3^2},\qquad  {\cal L}_2={\cal J}_1{\cal J}_2-\alpha\frac{s_1s_2}{s_3^2},\qquad
{\cal X}={\cal J}_3.\end{equation*}
The structure relations are
\begin{gather}\label{structure4}\{{\cal X},{\cal L}_1\}=-2{\cal L}_2,\qquad  \{{\cal X},{\cal L}_2\}=2{\cal L}_1-{\cal H}+{\cal X}^2+\alpha,\qquad
\{{\cal L}_1,{\cal L}_2\}=-2({\cal L}_1+\alpha){\cal X},\end{gather}
and the Casimir relation is
\begin{equation}\label{classcasimir}{\cal L}_1^2 + {\cal L}_2^2- {\cal L}_1 {\cal H}+ {\cal L}_1 {\cal X}^2+ \alpha{\cal X}^2+ \alpha{\cal L}_1=0.
\end{equation}
From the results of \cite{KALNINS} we know that additive separation of
variables in the Hamilton--Jacobi equation ${\cal H}=E$ is possible in
subgroup type coordinates in which ${\cal X}$, ${\cal L}_1$ or ${\cal
  S}=2({\cal L}_1-i{\cal L}_2)-{\cal H}+{\cal X}^2$, respectively, are
constants of separation. This corresponds to two choices of  spherical
coordinates and one of horospherical coordinates, respectively. We
seek two variable models for the Poisson bracket relations
\eqref{structure4}, \eqref{classcasimir}. There is also separation in
ellipsoidal coordinates (i.e., non-subgroup type coordinates) but we will not make use
of this here.

The justif\/ication for these models comes from Hamilton--Jacobi theory. The phase space for our problem is 4-dimensional. Thus it is possible to f\/ind canonical variables ${\cal H}$, ${\cal I}$, ${\cal Q}$, ${\cal P}$ such that
$\{{\cal I},{\cal H}\}=\{{\cal P},{\cal Q}\}=1$ and all other Poisson brackets vanish. In terms of $\cal H$ and the other canonical variables the Poisson bracket can be expressed  as
\begin{equation}\label{restrictedbracket1} \{{\cal F},{\cal G}\}= -\partial_{\cal H}{\cal F}\partial_{\cal I}{\cal G}+\partial_{\cal I}{\cal F}\partial_{\cal H}{\cal G}
-\partial_{\cal Q}{\cal F}\partial_{\cal P}{\cal G}+\partial_{\cal
  P}{\cal F}\partial_{\cal Q}{\cal G}.\end{equation}
(As follows from standard theory \cite{ARNOLD} one can construct a set of such
canonical variables from a complete integral of the Hamilton--Jacobi
equation. Our 2D second order superintegrable systems are always
multiseparable, and each separable solution of the Hamilton--Jacobi
equation provides a complete integral. Thus we can f\/ind these
canonical variables in several distinct ways.)
Now we restrict our attention to the algebra of constants of the motion. This algebra is generated by ${\cal H}$, ${\cal L}_1$, ${\cal L}_1$, ${\cal X}$, subject to the relation \eqref{classcasimir}. Thus, considered as functions of the canonical variables, the constants of the motion are independent of ${\cal H}$. If we further restrict the system to the constant energy space ${\cal H}=E$ then we can consider ${\cal H}$ as non varying  and every constant of the motion $\cal F$ can be expressed in the form ${\cal F}(E,{\cal Q},{\cal P})$. This means that the Poisson bracket of two constants of the motion, ${\cal F}$, ${\cal G}$ can be computed as
\[\{{\cal F},{\cal G}\}=-\partial_{\cal Q}{\cal F}\partial_{\cal P}{\cal G}+\partial_{\cal P}{\cal F}\partial_{\cal Q}{\cal G}.\]
Thus all functions depend on only two canonically conjugate variables ${\cal Q}$, ${\cal P}$ and the parameter~$E$. This shows the existence and the form of two variable models of conjugate variables. However the proof is not constructive and, furthermore, it is not unique. Two obtain constructive results we will use the strategy of  setting  ${\cal Q}$ equal to one of the constants of the motion that corresponds to separation of variables in some coordinate system, and then use~\eqref{restrictedbracket1} for the Poisson bracket and require that relations \eqref{structure4}, \eqref{classcasimir} hold.
In order to make clear that we are computing on the constant energy hypersurface expressed in canonical variables we will use a dif\/ferent notation.
We will set ${\cal Q}_E=c$, ${\cal P}_E=\beta$ so, ${\cal F}({\cal H},{\cal Q},{\cal P})=f(c,\beta)$, ${\cal G}({\cal H},{\cal Q},{\cal P})=g(c,\beta)$,  and
\[\{{\cal F},{\cal G}\}_E=\{f,g\}=-\partial_c{f}\partial_\beta{g}+\partial_\beta{f}\partial_c{g}.\]

For our f\/irst model we require $X\equiv {\cal X}_E=c$. Substituting this requirement  and ${\cal H}=E$ into the structure equations we obtain the result
\begin{alignat}{3}\label{model1}
 & I:\quad &&  L_1=\frac12(E-c^2-\alpha)+ \frac12\sqrt{c^4-2c^2(E+\alpha)+(E-\alpha)^2}\sin 2\beta,& \\
 &&& X=c,\qquad  L_2=\frac12\sqrt{c^4-2c^2(E+\alpha)+(E-\alpha)^2}\cos 2\beta.& \nonumber
 \end{alignat}
In this model, and in all other classical models, $\beta$ is not uniquely determined: we can replace it by $\beta'=\beta+k(c)$ for any function $k(c)$ and the variables $c$ and $\beta'$ remain canonically conjugate.

For a second model we require $L_1\equiv ({\cal L}_1)_E=c$ and proceed in a similar fashion. The result~is
\begin{alignat}{3} \label{model2} & II:\quad &&  L_1=c,\qquad  L_2=\sqrt{c(E-c-\alpha)}\sin (2\sqrt{c+\alpha}\beta),\\
&&& X=\sqrt{\frac{c(E-c-\alpha)}{c+\alpha}}\cos(2\sqrt{c+\alpha}\beta).& \nonumber\end{alignat}

For the third and last model we need to diagonalize the symmetry ${\cal S}=2({\cal L}_1-i{\cal L}_2)-{\cal H}+{\cal X}^2$ corresponding to separation in horospherical coordinates. For this it is convenient to rewrite the structure equations \eqref{structure4}, \eqref{classcasimir} in terms of the new basis ${\cal S}$, ${\cal L}_1+i{\cal L}_2$, ${\cal X}$:
\begin{gather*}%\label{structure5}
\{ {\cal S},{\cal X}, \}=2i({\cal S}+\alpha),\qquad  \{{\cal S},{\cal L}_1+i{\cal L}_2\}=-2i{\cal X}({\cal S}- 2{\cal X}^2+2{\cal H}+3\alpha),\\
 \{{\cal L}_1+i{\cal L}_2,{\cal X}\}=-i\left({\cal X}^2+2( {\cal L}_1+i{\cal L}_2)-{\cal H}+\alpha\right) .
\nonumber\end{gather*}
The Casimir relation is
\begin{gather*}%\label{classcasimir1}
-2{\cal S}({\cal L}_1 +i {\cal L}_2)-{\cal S}{\cal X}^2+{\cal X}^4+{\cal H}{\cal S}-2{\cal X}^2{\cal H}+{\cal H}^2
  -\alpha\left(2( {\cal L}_1 + i{\cal L}_2)+{\cal S}+3{\cal X}^2+ {\cal H}\right)=0.
\nonumber\end{gather*}
For model III we set $S=c$ and obtain
\begin{alignat*}{3} %\label{model3}
 & III:\quad &&  S=c,\qquad X=-2i(c+\alpha)\beta,& \\
&&&  L_1+iL_2=8(c+\alpha)^3\beta^4+2(c+\alpha)(3\alpha+c+2E)\beta^2-\frac{(c+E)(\alpha-E)}{2(c+\alpha)}.&
\nonumber\end{alignat*}

\subsection[Classical model $\to$ quantum model]{Classical model $\boldsymbol{\to}$ quantum model}

What have we achieved with these classical models? For one thing they
show us how to paramete\-rize the constants of the motion and exhibit
their functional dependence. More important for our purposes, they
give us a rational means to derive the possible one-variable quantum
mo\-dels. This may seem surprising. How can classical mechanics
determine quantum mechanics uniquely? How can structures such as the
Wilson family of orthogonal polynomials, contai\-ning the Hahn
polynomials, be derived directly from classical mechanics? The point
is that the structures we are studying are second order
superintegrable systems in 2D. In papers \cite{KKMW,Kress2007,KKM20061} it has been shown
that there is a 1-1 relationship between the quantum and classical
versions for such systems, for all 2D Riemannian spaces. (Similarly
there is a 1-1 relationship in
3D for nondegenerate potentials on conformally f\/lat spaces.)  The
structures are not identical, since as we can see from the examples in
this paper, the structure relations in the classical and quantum cases
are not identical; there are quantum modif\/ications of the classical
equations. Although we know of no direct prescription for their
determination, nonetheless the quantum structure equations  are uniquely determined by the
classical structure equations. Given a second system of second order
constants of the motion we write down the corresponding quantum system
via the usual correspondence, where products of classical functions
are replaced by symmetrized quantum operators, and generate the
quadratic algebra by taking repeated commutators. Even order classical
symmetries correspond to formally self-adjoint quantum symmetries, and
odd order classical symmetries correspond to formally skew-adjoint
quantum symmetries. (This relationship no longer holds for third order
superintegrable systems~\cite{GW,GRAVEL}.) We will demonstrate here how to get
quantum models from the classical ones that we have derived.

The basic prescription for the transition from the classical case to the operator case  is to replace a pair of canonically conjugate variables
$c$, $\beta$
by $c\to t$, $\beta\to \partial_t$. (There is no obstruction to quantization for second order superintegrable systems.) Once an appropriate
choice of $\beta$ is made in a classical model, we can use this
prescription to go to a dif\/ferential operator model of the quantum structure
equations.
In particular model III above suggests a operator  model such that $S$
is multiplication by $c$, $X$ is a f\/irst order dif\/ferential operator
in $c$
and $L_1+iL_2$ is a fourth order dif\/ferential operator. The result,
whose existence is implied by the 1-1 classical/quantum relationship
for second order superintegrable systems, is
\begin{alignat}{3} \label{model3q} & III: \quad &&  S=t,\qquad X=-2i(t+\alpha)\partial_t+2i,& \\
&&& L_1+iL_2=8(t+\alpha)^3\partial_t^4+2(t+\alpha)(3\alpha+t+2E+9)\partial_t^2-2(t+5\alpha+4E+18)\partial_t&
\nonumber\\
&&& \phantom{L_1+iL_2=}{} +\left(2+\frac{E}{2}-\frac{\alpha}{2}\right)+\frac{E^2+2E(9-\alpha)+(\alpha+12)(\alpha+6)}{2(t+\alpha)}.&
\nonumber\end{alignat}
The leading order dif\/ferential operators terms agree with the
classical case but there are lower order correction terms needed to
correct for the noncommutivity of $t$ and $\partial_t$. We can realize
various irreducible representations of the quadratic algebra by
choosing subspaces of functions of $t$ on which the operators
act. This model agrees with~\eqref{Ldef1}, \eqref{bddbelow1}, \eqref{kappa0}
in the case where $C(n-1,n)=1$ and $C(n+1,n)$ is fourth order. However, there we had a space spanned by a~countable number of eigenvectors of the skew-adjoint symmetry $X$ whereas here we want the spectral decomposition of the self-adjoint symmetry $S$ to
govern the model. This forces $L_2$ to be skew-adjoint and $X$ to be self-adjoint. Thus, though the dif\/ferential operators are
formally the same, the Hilbert spaces and the spectral analysis are dif\/ferent. All the representations are inf\/inite-dimensional. One class can be realized by closing the dense subspace  of $C^\infty$ functions with compact support on $0<t<\infty$  where the measure is $dt/t$. The  the  spectrum of $S$ is continuous and runs over the positive real axis. Here $X$ also has continuous real spectra covering the full  real axis. In particular the generalized eigenfunction of $X$ with real eigenvalue~$\lambda$ is proportional to $t^{-i\lambda}$, and $\mu$ is pure imaginary. Thus the spectral analysis of $X$ is given by the Mellin transform. There is a similar irreducible representation def\/ined on $-\infty<t<0$. By a canonical transformation we can also get models of these representations in which both $C(n-1,n)=1$ and $C(n+1,n)$ are second order. (We shall illustrate this explicitly for model~I.) Then the spectral decomposition of $S$ is given by the Hankel transform. Since these particular eigenspaces of $H$ admit no discrete spectrum for any of the symmetries of interest, we shall not analyze them further.

Now we consider model I, \eqref{model1}. Due to the presence of
trigonometric terms in $\beta$ we cannot realize this  as a f\/inite
order dif\/ferential operator model. However, we can perform a hodograph
transformation, i.e.\ use the prescription $\beta\to t$, $c\to
-\partial_t$
to realize the model. This would seem to make no sense due to the
appearance of functions of $c$ under the square root sign. However,
before using the prescription we can make use of the freedom to make a
replacement $\beta'=\beta +g(c)$ which preserves canonical variables.
We choose
\[e^{-2i\beta}\rightarrow
e^{-2i\beta}/\sqrt{c^4-2c^2(E+\alpha)+(E-\alpha)^2}
\]
but leave $c$ unchanged. Then we f\/ind
\begin{gather} L_1=\frac12(E-c^2-\alpha)-\frac{i}{4}\left[\left(c^4-2c^2(E+\alpha)+(E-\alpha)^2\right)e^{2i\beta}-1\right],
\nonumber\\
L_2=-\frac{i}{4}\left[\left(c^4-2c^2(E+\alpha)+(E-\alpha)^2\right)e^{2i\beta}+1\right],\label{model3qq}
\end{gather}
with $X$ as before. Now we apply the quantization prescription
$\beta\to t$, $c\to -\partial_t$ and obtain a~model in which both $L_1$
and $L_2$ are fourth order and $X$ is a f\/irst order dif\/ferential
operator. This is, in fact, identical to within a coordinate change to
model \eqref{model3q}. One might also try to obtain a dif\/ference
operator model from \eqref{model3qq} with the replacement $c\to
  t$, $\beta\to \partial_t$, so that~$e^{2i\beta}$ would become a
  dif\/ference operator. However, this dif\/ference operator quantum model is equivalent to
  what would get from the $\beta\to t$, $c\to -\partial_t$ model by
  taking a Fourier transform. Thus we don't regard it as new.

There is an alternate way to obtain a quantum realization from model
I. We use the fact that
\[c^4-2c^2(E+\alpha)+(E-\alpha)^2=(c^2-(E+\alpha))^2-4\alpha\]
and set
\[\phi=\arctan \left(\frac{\sqrt{-4\alpha}}{c^2-(E+\alpha)^2}\right).\]
Now we let $2\beta\to 2\beta+\phi$ to obtain
\begin{gather*}
L_1=\frac12(E-c^2-\alpha)+\frac12\left((c^2-(E+\alpha)^2)\sin
2\beta+2i\sqrt{\alpha}\cos 2\beta\right),\\
L_2=\frac12\left((c^2-(E+\alpha)^2)\cos
2\beta-2i\sqrt{\alpha}\sin 2\beta\right),\qquad X=c.
\end{gather*}
Now the prescription $\beta\to t$, $c\to -\partial_t$ leads to a quantum
realization of $L_1$, $L_2$ by second order dif\/ferential
operators. Indeed
\begin{gather*}
 L_1=\frac12(\cos(2t) -1)\partial_t^2-8i\xi
\sin(2t)\partial_t+\left(-\frac{E}{2}+64\xi^2+8i\xi-\frac14-\frac{\alpha}{2}\right)\cos(2t)+\frac{E-\alpha}{2},
\\ L_2=\frac12\sin(2t)\partial_t^2+8i\xi
\cos(2t)\partial_t+\left(-\frac{E}{2}+64\xi^2+8i\xi-\frac14-\frac{\alpha}{2})\cos(2t\right)+\frac{E-\alpha}{2},\\
 X=\partial_t.
\end{gather*}
Here $\xi$ is arbitrary and can be removed via a gauge
transformation. The change of variable \mbox{$\tau=e^{2it}$} reduces this
model to the form~\eqref{model0}. This shows that the f\/lexibility we
had in construc\-ting dif\/ferential operator models from the abstract
representation theory by renormalizing our basis vectors $f_n$ is
replaced in the classical model case by appropriate canonical
transformations $c\to c$, $\beta\to \beta+g(c)$. In either case there
is essentially only one dif\/ferential operator model that can be
transformed in various ways.

It is clear that model II cannot produce f\/inite order dif\/ferential operator realizations of the quantum quadratic algebra, due to the intertwining of square root dependence for $c$ and exponential dependence for $\beta$. However, it will produce a dif\/ference operator realization via Taylor's theorem: $e^{a\partial_t}f(t)=f(t+a)$.To show this explicitly we make a coordinate change such that $2\sqrt{c+\alpha}\partial_c=\partial_{\cal C}$ in \eqref{model2}, which suggests realizations of the quantum operators in the form
\begin{gather} L_1f(t)=(t^2-\alpha)f(t),\qquad Xf(t)=h(t)f(t+i)+m(t)f(t-i),\nonumber\\
L_2f(t)=- \frac{i}{2}(i+2t)h(t)f(t+i)+\frac{i}{2}(-i+2t)m(t)f(t-i).\label{model2q}\end{gather}
A straightforward computation shows that the quantum algebra structure equations are satisf\/ied if and only if
\begin{equation}\label{model2q1} h(t)m(t+i)=\frac14\frac{(\alpha-t^2-it)(t^2+it-E)}{t(t+i)}.\end{equation}
Since $\alpha=-a^2+\frac14$ and $E=-(\mu-1+a)^2+\frac14$ for bounded below representations, we can factor~\eqref{model2q1} simply to obtain
\begin{gather} h(t)m(t+i)=-\frac{1}{4t(t+i)} \left(t+\frac{i}{2}+ia\right)\left(t+\frac{i}{2}-ia\right)\nonumber\\
\phantom{h(t)m(t+i)=}{}\times \left(t+\frac{i}{2}+i\mu+ia\right)\left(t+\frac{3i}{2}-i\mu-ia\right).\label{model2q2}
\end{gather}
Note that only the product \eqref{model2q1} is determined, not the individual factors. Thus we can choose~$h(t)$, say, as an arbitrary nonzero function and then determine $m(t)$ from \eqref{model2q1}. All these modif\/ications of the factors are accomplished by gauge transformations on the representation space: ${\tilde f}(t)=\rho(t)f(t)$ where $\rho(t)$ is the gauge function. If we choose the factors in the form
\[h(t)=i \frac{(\frac{1}{2}-a-it)(\mu+a-\frac{1}{2}-it)}{2t},\qquad m(t)=
-i\frac{(\frac{1}{2}-a+it)(\mu+a-\frac{1}{2}+it)}{2t},\]
then we we get exactly the model~\eqref{difference2}. The f\/inite
dimensional model is related by the simple change of variables
$t\to i(t-a+1/2)$, $\mu=-m  $. In any case, there is only a single solution of these equations, up to a gauge transformation.

\subsection{The classical model for S9}

This is the system on the complex sphere, with nondegenerate potential
\[
V=\frac{a_1}{s_1^2}+\frac{a_2}{s_2^2}+\frac{a_3}{s_3^2},
\]
where $s_1^2+s_2^2+s_3^2=1$.
The classical S9 system has a basis of symmetries
\begin{equation}\label{comS9} {\cal L}_1={\cal
  J}_1^2+a_2\frac{s_3^2}{s_2^2}+a_3\frac{s_2^2}{s_3^2},\qquad {\cal L}_2={\cal
  J}_2^2+a_3\frac{s_1^2}{s_3^2}+a_1\frac{s_3^2}{s_1^2},\qquad
{\cal L}_3={\cal
  J}_3^2+a_1\frac{s_2^2}{s_1^2}+a_2\frac{s_1^2}{s_2^2},
\end{equation}
where ${\cal H}={\cal L}_1+{\cal L}_2+{\cal L}_3+a_1+a_2+a_3$
and the ${\cal J}_i$ are def\/ined by ${\cal J}_3=s_1p_{s_2}-s_2p_{s_1}$
and cyclic permutation of indices. The classical structure relations
are
\begin{gather*}%\label{S9classstruc}
\{{\cal L}_1,{\cal R}\}=8{\cal L}_1({\cal H}+a_1+a_2+a_3)-8{\cal
  L}_1^2-16{\cal L}_1{\cal L}_2-16a_2{\cal L}_2\\
  \phantom{\{{\cal L}_1,{\cal R}\}=}{} +16a_3({\cal
  H}+a_1+a_2+a_3-{\cal L}_1-{\cal L}_2),\nonumber
\\
\{{\cal L}_2,{\cal R}\}=-8{\cal L}_2({\cal H}+a_1+a_2+a_3)+8{\cal
  L}_2^2+16{\cal L}_1{\cal L}_2+16a_1{\cal L}_1\nonumber\\
  \phantom{\{{\cal L}_2,{\cal R}\}=}{} -16a_3({\cal
  H}+a_1+a_2+a_3-{\cal L}_1-{\cal L}_2),
\nonumber\end{gather*}
with $\{{\cal L}_1,{\cal L}_2\}={\cal R}$ and
\begin{gather*}
{\cal R}^2-16{\cal L}_1{\cal L}_2({\cal H}+a_1+a_2+a_3) +16{\cal
  L}_1^2{\cal L}_2+16{\cal L}_1{\cal L}_2^2+16a_1{\cal
  L}_1^2\\
  \qquad{} +16a_2{\cal L}_2^2+16a_3({\cal H}+a_1+a_2+a_3)^2
-32a_3({\cal
  H}+a_1+a_2+a_3)({\cal L}_1+{\cal L}_2)\\
  \qquad{} +16a_3{\cal L}_1^2+32a_3{\cal
  L}_1{\cal L}_2+16a_3{\cal L}_2^2-64a_1a_2a_3=0.
\end{gather*}
Taking ${\cal L}_1=c$, ${\cal H}=E$ with $c$, $\beta$ as conjugate variables, we f\/ind the model
\begin{gather}
{\cal L}_2=\frac12(a_1+2a_2+E-c)-\frac{(a_2-a_3)(a_1+2a_2+2a_3+E)}{2(c+a_2+a_3)}
\nonumber  \\
{}+ \frac{\sqrt{(4a_1a_2\,{+}\,4a_1a_3\,{+}\,2c(E\,{+}\,a_1\,{+}\,a_2\,{+}\,a_3)\,{+}\,
 4ca_1\,{-}\,(E\,{+}\,a_1\,{+}\,a_2\,{+}\,a_3)^2\,{-}\,c^2)(4a_2a_3\,{-}\,c^2)}}
 {2(a_2\,{+}\,a_3\,{+}\,c)}\nonumber\\
{}\times \cos(4\beta\sqrt{a_2+a_3+c}).\label{S9clssmodel}
\end{gather}
This suggests a dif\/ference operator realization of the quantum
model.

In the quantum case the symmetry operators $L_1$, $L_2$, $L_3$ are obtained from the
corresponding classical constants of the motion \eqref{comS9}
through the replacements ${\cal J}_k\to J_k$ where the angular
momentum operators $J_k$ are def\/ined by
$J_3=x_1\partial_{x_2}-x_2\partial_{x_1}$
and cyclic permutation of indices. Here  $ H= L_1+ L_2+ L_3+a_1+a_2+a_3$.
 The quantum structure relations
can be put in the symmetric form
\begin{gather*}%\label{S9quantstruc1}
[L_i,R]=4\{L_i,L_k\}-4\{L_i,L_j\}- (8+16a_j)L_j + (8+16a_k)L_k+ 8(a_j-a_k),\\
%\label{S9quantstruc2}
R^2=\frac86\{L_1,L_2,L_3\}+ -(16a_1+12)L_1^2 -(16a_2+12)L_2^2  -(16a_3+12)L_3^2\\
\phantom{R^2=}{}+\frac{52}{3}(\{L_1,L_2\}+\{L_2,L_3\}+\{L_3,L_1\})+ \frac13(16+176a_1)L_1\nonumber\\
\phantom{R^2=}{}+\frac13(16+176a_2)L_2 + \frac13(16+176a_3)L_3 +\frac{32}{3}(a_1+a_2+a_3)\nonumber\\
\phantom{R^2=}{}+48(a_1a_2+a_2a_3+a_3a_1)+64a_1a_2a_3.\nonumber
\end{gather*}
Here $i$, $j$, $k$ are chosen such that $\epsilon_{ijk}=1$ where $\epsilon$
is the pure skew-symmetric tensor,  $R=[L_1,L_2]$ and
$\{L_1,L_j\}=L_iL_j+L_jL_i$ with an analogous def\/inition of
$\{L_1,L_2,L_3\}$ as a sum of~6 terms.
In practice we will substitute $L_3=H-L_1-L_2-a_1-a_2-a_3$ into these equations.

Proceeding exactly as in the S3 case \eqref{model2q},
\eqref{model2q1}, \eqref{model2q2}, we f\/ind that  the dif\/ference operator analogy of \eqref{S9clssmodel} for the quantum quadratic algebra is
\begin{gather*}%\label{S9quantmodel}
L_1=4t^2-\frac12+\beta^2+\gamma^2,\\
L_2=h(t)T^i+m(t)T^{-i}+\ell(t)\nonumber\\
{}=\frac{[-4\alpha^2-8\alpha-4+4{\cal E}^2+16i(\alpha+1)t+16t^2](\beta+1+\gamma-2it)(\beta-1-\gamma+2it)}{1024t(t+i)(2t+i)^2}\nonumber\\
{}\times [-4\alpha^2-4+8\alpha+4{\cal E}^2+16i(1-\alpha)t+16t^2](\beta+1-\gamma-2it)(\beta-1+\gamma+2it)T^i +T^{-i}\nonumber\\
{}+\left[-2t^2-\frac12{\cal E}^2-\frac12\beta^2+\frac12\alpha^2+\frac12\gamma^2+\frac{(\gamma^2-\beta^2)
(-4\alpha^2+4{\cal E}^2)}{8(1+4t^2)}\right],
\nonumber
\end{gather*}
where
\[a_1=\frac14-\alpha^2,\qquad a_2=\frac14-\beta^2,\qquad a_3=\frac14-\gamma^2,\qquad
H=\frac14-{\cal E}^2.
\]
The quadratic terms factor into simple linear terms, and just as in
the S3 case, it is only~$\ell(t)$ and the product $h(t)m(t+i)$ that is
uniquely determined. We can change the individual factors~$h(t)$,~$m(t)$ by a gauge
transformation.  With the change of variable $t=i\tau$  and a gauge
transformation to an operator with maximal symmetry in $\tau$,  we obtain the standard
model
\begin{gather*} h(t)={\tilde
  h}(\tau)= \frac{(A+\tau)(B+\tau)(C+\tau)(D+\tau)}{4\tau(\tau+1/2)},
\\
 m(t)={\tilde m}(\tau)=
\frac{(A-\tau)(B-\tau)(C-\tau)(D-\tau)}{4\tau(\tau-1/2)},
\\
A=\frac{{\cal E}+\alpha+1}{2},\qquad B=\frac{{\cal E}-\alpha+1}{2},\qquad
C=\frac{\beta+\gamma+1}{2},\qquad D=\frac{\beta-\gamma+1}{2}.
\end{gather*}
It follows that $L_2= {\tilde h}(\tau)E^{+1}+{\tilde m}(\tau)E^{-1}+{\tilde \ell}(\tau)$ is a linear combination of $L_1$ and the
dif\/ference operator whose eigenfunctions are the Wilson polynomials,
just as found in~\cite{KMPost}. Here $E^sf(\tau)=f(\tau +s)$.

\section{Conclusions and prospects}
This paper consists of two related parts. In the f\/irst part we have studied the representation theory for the quadratic algebra associated with a 2D second order quantum superintegrable system with degenerate potential, namely S3. We have classif\/ied the possible f\/inite-dimensional representations and inf\/inite dimensional bounded below representations, i.e., those with a lowest weight vector. Then we have constructed the possible Hilbert space models for these representations, in terms of dif\/ferential operators or of dif\/ference operators acting on spaces of functions of one complex variable. These models make it easy to f\/ind raising and lowering operators for the representations and to uncover relationships between the algebras and families of orthogonal polynomials. Here S3 has been treated as an example of a degenerate potential superintegrable system. The example S9 of a nondegenerate potential was treated in \cite{KMPost}. In 2D there are 13~equivalence classes of superintegrable systems with nontrivial potentials: 7 nondegenerate and 6 degenerate. Results for all of these cases will be included in the thesis of the third author.

In the  second part of this work we have taken up the study of models
of the quadratic algebras associated with the classical second order
superintegrable systems. In each  model there is only a single pair of canonically conjugate variables, rather than the 2 pairs in the original classical system. We showed, based on classical Hamilton--Jacobi theory, that such models always exist. Then we described a procedure to derive the one variable models for the quantum quadratic algebras from the models for the classical quadratic algebras. Since it is known that there is a 1-1 relationship between classical and quantum second order superintegrable systems (even though the algebras are not the same), it is not too surprising that one should be able to compute the quantum models from the classical models. However, we have made this explicit.
We applied this procedure not only to obtain the dif\/ferential and dif\/ference operator models for system S3, but also for the generic system S9. For S9 we showed that there is a dif\/ference operator model associated with general Wilson polynomials, but no dif\/ferential operator model. This construction demonstrates that the theory of general Wilson polynomials is imbedded in classical mechanics in a manner quite dif\/ferent from the usual group theory (Racah polynomial)  approach.

There is much more work to be done. Once the models are worked out and the corresponding
  functional Hilbert spaces are constructed, usually Hilbert spaces
  with kernel function, then one needs to f\/ind intertwining operators
  that map the model space to the space on which the quantum
  Schr\"odinger operator is def\/ined. Also,we have demonstrated
  how to determine the classical models and show how they quantize in a
  unique fashion. A puzzle here is that we are f\/inding classical
  models corresponding to non-hypergeometric type variable
  separation. These classical models typically involve elliptic
  functions. We do not yet understand how they can be quantized. They
  clearly do not lead to dif\/ferential or ordinary dif\/ference operator
  quantum models.

Another part of our ef\/fort  is to study the structure of quadratic algebras
  corresponding to 3D nondegenerate superintegrable systems, and to f\/ind
  two variable models for them. This is a much more dif\/f\/icult problem
  than in 2D, where it led to general Wilson and Racah polynomials,
  among other models. The quadratic algebra still closes at order 6
  but now there are 6 linearly second order symmetries, rather than 3,
  and they are functionally dependent, satisfying a polynomial
  relation of order 8. There are 4 commutators, instead of 1.  For the
  models we expect to f\/ind multivariable extensions of Wilson
  polynomials, among many other constructs.

\pdfbookmark[1]{References}{ref}
\LastPageEnding


\begin{thebibliography}{99}

\footnotesize\itemsep=0pt

%-----------------------------------------------------------------------
\bibitem{WOJ}
Wojciechowski S.,
Superintegrability of the Calogero--Moser system,
{\it Phys. Lett.~A} {\bf 95} (1983),  279--281.
%-----------------------------------------------------------------------
\bibitem{EVA}Evans N.W.,
Superintegrability in classical mechanics,
{\it Phys. Rev.~A} {\bf 41} 1990, 5666--5676.\\
Evans N.W., Group theory of the
Smorodinsky--Winternitz system, {\it J. Math. Phys.} {\bf 32} (1991), 3369--3375.
%-----------------------------------------------------------------------
\bibitem{EVAN}
Evans N.W.,
Super-integrability of the Winternitz system,
{\it Phys. Lett.~A} {\bf 147} (1990), 483--486.
%-----------------------------------------------------------------------
\bibitem{FMSUW}
Fri\v s J., Mandrosov V., Smorodinsky Ya.A.,  Uhl\'\i r  M., Winternitz P.,
On higher symmetries in quantum mechanics,
{\it Phys. Lett.}  {\bf 16} (1965), 354--356.
%-----------------------------------------------------------------------

\bibitem{BDK}
Bonatos D., Daskaloyannis C., Kokkotas K.,
Deformed
oscillator algebras for two-dimensional quantum superintegrable
systems,
{\it Phys. Rev.~A} {\bf 50} (1994), 3700--3709, \href{http://arxiv.org/abs/hep-th/9309088}{hep-th/9309088}.
%-----------------------------------------------------------------------
\bibitem{CDas}
Daskaloyannis C.,
Quadratic Poisson algebras of two-dimensional classical
superintegrable systems and quadratic associative algebras of quantum
superintegrable systems,
{\it J. Math. Phys.} {\bf 42} (2001), 1100--1119, \href{http://arxiv.org/abs/math-ph/0003017}{math-ph/0003017}.
%-----------------------------------------------------------------------

\bibitem{VILE}
Letourneau  P., Vinet L.,
Superintegrable systems:
polynomial algebras and quasi-exactly solvable
Hamiltonians,
{\it Ann. Phys.} {\bf 243} (1995), 144--168.
%-----------------------------------------------------------------------
\bibitem{RAN}
Ra\~nada M.F.,
Superintegrable $n=2$ systems, quadratic constants of motion, and
potentials of Drach,
{\it J.~Math. Phys.} {\bf 38} (1997), 4165--4178.
%-----------------------------------------------------------------------
\bibitem{KMWP}
Kalnins E.G., Miller W.~Jr.,  Williams G.C., Pogosyan G.S.,
On superintegrable symmetry-breaking potentials in
  $n$-dimensional Euclidean space,
{\it J.~Phys.~A:~Math.~Gen.} {\bf 35} (2002), 4655--4720.
% 4705--4720
%-----------------------------------------------------------------------
\bibitem{KMJP2}
Kalnins E.G., Miller W. Jr.,  Pogosyan G.S.,
Completeness of multiseparable superintegrability in $E_{2,C}$,
{\it J.~Phys.~A:~Math.~Gen.} {\bf 33}  (2000), 4105--4120.
%-----------------------------------------------------------------------
\bibitem{KMJP3}
Kalnins E.G., Miller W.\ Jr., Pogosyan G.S.,
Completeness of multiseparable superintegrability on the complex 2-sphere,
{\it J.~Phys.~A:~Math.~Gen.} {\bf 33} (2000), 6791--6806.
%-----------------------------------------------------------------------
\bibitem{Koenigs}
Koenigs G., Sur les g\'eod\'esiques a int\'egrales quadratiques, A note
appearing in ``Lecons sur la th\'eorie g\'en\'erale des
surfaces'', G. Darboux, Vol.~4, Chelsea Publishing, 1972, 368--404.
%-----------------------------------------------------------------------

\bibitem{CAL}
Calogero F.,
Solution to the one-dimensional $N$-body problems with quadratic
and/or inversely quadratic pair potentials,
{\it J.  Math. Phys.} {\bf  12} (1971),  419--436.
%-----------------------------------------------------------------------

\bibitem{WOJW}
Rauch-Wojciechowski  S., Waksj\"o C.,
What an ef\/fective criterion of separability says about the Calogero
type systems,
{\it J. Nonlinear Math. Phys.} {\bf  12} (2005), suppl.~1, 535--547.
%-----------------------------------------------------------------------

\bibitem{KKMP}
Kalnins E.G., Kress J.M., Miller W.~Jr., Pogosyan G.S.,
Completeness of superintegrability in two-dimensional constant
curvature spaces,
{\it J.~Phys.~A:~Math.~Gen.}  {\bf 34} (2001), 4705--4720, \href{http://arxiv.org/abs/math-ph/0102006}{math-ph/0102006}.
% 4705--4720
%-----------------------------------------------------------------------
\bibitem{Zhedanov1991}Zhedanov A.S., ``Hidden symmetry'' of
  Askey--Wilson polynomials, {\it Theoret. and Math. Phys.} {\bf 89} (1991),
1146--1157.
%-----------------------------------------------------------------------
\bibitem{Zhedanov1992a} Granovskii Ya.I., Zhedanov A.S., Lutsenko  I.M., Quadratic algebras and dynamics in curved spaces. I.~Oscillator,
{\it Theoret. and Math. Phys.} {\bf 89}  (1992), 474--480.
%-----------------------------------------------------------------------
\bibitem{Zhedanov1992b} Granovskii Ya.I., Zhedanov A.S., Lutsenko I.M., Quadratic algebras and dynamics in curved spaces. II.~The
Kepler problem, {\it Theoret. and Math. Phys.} {\bf 91} (1992), 604-612.
%-----------------------------------------------------------------------
%\bibitem{Dask2001}Daskaloyannis C., Quadratic Poisson algebras of two-dimensional classical superintegrable systems and
%quadratic associative algebras of quantum superintegrable systems, {\it J. Math. Phys}, (2001), {\bf 42},  1100-1119,
%math-ph/0003017.
%-----------------------------------------------------------------------
\bibitem{Quesne1994} Quesne C., Generalized deformed parafermions, nonlinear deformations of so(3) and exactly solvable potentials,
{\it  Phys. Lett.~A} {\bf  193} (1994), 245--250.
%-----------------------------------------------------------------------


\bibitem{TTW}
Tempesta P.,  Turbiner A.V., Winternitz P., Exact solvability of
superintegrable systems,
{\it J. Math. Phys.} {\bf 42} (2001), 4248--4257.
%-----------------------------------------------------------------------
\bibitem{GW}  Gravel S., Winternitz P., 	Superintegrability
  with third-order integrals in quantum and classical mechanics,
{\it J. Math. Phys.}  {\bf 43} (2002),  5902--5912, \href{http://arxiv.org/abs/math-ph/0206046}{math-ph/0206046}.
%-----------------------------------------------------------------------
\bibitem{HER1}
Ballesteros A., Herranz F., Santander M., Sanz-Gil T.,
Maximal superintegrability on $N$-dimensional
curved spaces,  {\it J.~Phys.~A: Math. Gen.} {\bf 36} (2003), L93--L99, \href{http://arxiv.org/abs/math-ph/0211012}{math-ph/0211012}.
%-----------------------------------------------------------------------
\bibitem{KKM20041}
Kalnins E.G., Kress J.M., Miller W.~Jr.,
Second  order superintegrable systems in conformally
f\/lat spaces.  I.~2D classical structure theory, {\it J. Math. Phys.} {\bf 46} (2005),
 053509, 28~pages.
%-----------------------------------------------------------------------
\bibitem{KKM20042}
Kalnins E.G., Kress J.M.,  Miller W.~Jr.,
Second  order superintegrable systems in conformally
f\/lat spaces.  II.~The classical 2D St\"ackel transform, {\it J.~Math. Phys.} {\bf 46}
(2005), 053510, 15~pages.
%-----------------------------------------------------------------------
\bibitem{KKM20051}
Kalnins E.G., Kress J.M.,  Miller W.~Jr.,
Second  order superintegrable systems in conformally
f\/lat spaces.  III.~3D  classical structure theory, {\it
  J. Math. Phys.} {\bf 46} (2005), 103507, 28 pages.
%-----------------------------------------------------------------------
\bibitem{KKM20052}
Kalnins E.G., Kress J.M.,  Miller W.~Jr.,
Second  order superintegrable systems in conformally
f\/lat spaces.  IV.~The classical 3D St\"ackel transform and 3D
classif\/ication theory, {\it
  J. Math. Phys.} {\bf  47} (2006), 043514, 26~pages.
%-----------------------------------------------------------------------
\bibitem{KKM20061}
Kalnins E.G., Kress J.M., Miller W.~Jr.,
Second  order superintegrable systems in conformally
f\/lat spaces.  V.~2D and 3D quantum systems, {\it
  J. Math. Phys.} {\bf 47} (2006),  093501, 25 pages.
%-----------------------------------------------------------------------
\bibitem{KMP2005}
Kalnins E.G.,  Miller W.~Jr., Pogosyan G.S.,
Exact and quasi-exact solvability of second order superintegrable
systems.  I.~Euclidean space preliminaries, {\it J. Math. Phys.} {\bf
  47} (2006), 033502, 30~pages, \mbox{\href{http://arxiv.org/abs/math-ph/0412035}{math-ph/0412035}}.
%-----------------------------------------------------------------------
\bibitem{KKW}
Kalnins E.G., Kress J.M., Winternitz  P.,
Superintegrability in a two-dimensional space of non-constant curvature,
{\it J.~Math.~Phys.} {\bf 43} (2002),  970--983, \href{http://arxiv.org/abs/math-ph/0108015}{math-ph/0108015}.

%-----------------------------------------------------------------------
\bibitem{KKMW}
Kalnins E.G., Kress J.M., Miller W.~Jr., Winternitz P.,
 Superintegrable systems in Darboux spaces,
{\it J.~Math.~Phys.} {\bf 44} (2003),  5811--5848, \href{http://arxiv.org/abs/math-ph/0307039}{math-ph/0307039}.
%
%-----------------------------------------------------------------------

\bibitem{DASK2005}
Daskaloyannis C., Ypsilantis K.,          Unif\/ied treatment and classif\/ication of  superintegrable
systems with integrals quadratic in momenta
on a two dimensional manifold, {\it J. Math. Phys.} {\bf 47}  (2006),
042904, 38~pages, \href{http://arxiv.org/abs/math-ph/0412055}{math-ph/0412055}.
%-----------------------------------------------------------------------
\bibitem{HMS}
 Horwood J.T., McLenaghan R.G., Smirnov R.G.,
Invariant classif\/ication of orthogonally separable Hamiltonian systems
in Euclidean space,
{\it Comm.  Math. Phys.} {\bf  259} (2005),  679--709, \href{http://arxiv.org/abs/math-ph/0605023}{math-ph/0605023}.
%-----------------------------------------------------------------------


\bibitem{SCQS} Tempesta P., Winternitz P., Miller W., Pogosyan G. (Editors),  Superintegrability in classical
and quantum systems, {\it CRM Proceedings Lecture Notes}, Vol.~37, American Mathematical Society, Providence, RI, 2004.
%-----------------------------------------------------------------------
\bibitem{KKM2007a}
Kalnins E.G., Kress J.M., Miller W.~Jr.,
Nondegenerate 2D complex Euclidean superintegrable systems and
algebraic varieties, {\it J. Phys. A: Math. Theor.}  {\bf  40}  (2007),  3399--3411.
 %-----------------------------------------------------------------------

\bibitem{KKM2007b}Kalnins E.G., Kress J.M., Miller W.~Jr.,
Fine structure for 3D second order superintegrable systems:
3-para\-meter potentials, {\it  J. Phys. A: Math. Theor.} {\bf 40}
(2007), 5875--5892.
%-----------------------------------------------------------------------

\bibitem{KKM2007c}
Kalnins E.G., Kress J.M., Miller W.~Jr.,
   Nondegenerate 3D complex Euclidean superintegrable systems and algebraic varieties, {\it J. Math. Phys.} {\bf 48} (2007), 113518,  26~pages, \href{http://arxiv.org/abs/0708.3044}{arXiv:0708.3044}.
%-----------------------------------------------------------------------
\bibitem{Quesne2007}
Quesne C.,
Quadratic algebra approach to an exactly solvable
position-dependent mass Schr\"odinger equation
in two dimensions,
{\it SIGMA} {\bf  3}  (2007), 067, 14~pages, \href{http://arxiv.org/abs/0705.2577}{arXiv:0705.2577}.
%-----------------------------------------------------------------------
\bibitem{DASK2007}
Daskaloyannis C., Tanoudis Y.,         Quantum   superintegrable
systems with quadratic integrals
on a two dimensional manifold, {\it J. Math. Phys.} {\bf 48} (2007), 072108, 22 pages, \href{http://arxiv.org/abs/math-ph/0607058}{math-ph/0607058}.
%-----------------------------------------------------------------------
\bibitem{KMPost}Kalnins E.G., Miller W. Jr., Post S., Wilson
  polynomials and the generic superintegrable system on the 2-sphere,
  {\it  J. Phys. A: Math. Theor.} {\bf 40} (2007), 11525--11538.
%-----------------------------------------------------------------------
\bibitem{GRAVEL}
Gravel S.,
Hamilton separable in Cartesian coordinates and third-order
integrals of motion,
{\it J.\ Math.\ Phys.} {\bf 45} (2004), 1003--1019.
%-----------------------------------------------------------------------

\bibitem{BH}
Ballesteros A., Herranz F.J.,
Universal integrals for superintegrable systems on $N$-dimensional
spaces of constant curvature, {\it J. Phys. A: Math. Theor.} {\bf
  40} (2007), F51--F59, \href{http://arxiv.org/abs/math-ph/0610040}{math-ph/0610040}.
%-----------------------------------------------------------------------

\bibitem{FORDY} Fordy A.P.,   Quantum super-integrable systems as
  exactly solvable models,  {\it SIGMA}  {\bf 3}  (2007),  025, 10~pages, \href{http://arxiv.org/abs/math-ph/0702048}{math-ph/0702048}.
%-----------------------------------------------------------------------


\bibitem{KOP2}Kalnins E.G., Miller W.\ Jr., Pogosyan G.S., Exact
  and quasi-exact solvability of second order superintegrable quantum
  systems. II. Connection with separation of variables, {\it
  J. Math. Phys.} {\bf 48} (2007), 023503, 20 pages.
%-----------------------------------------------------------------------
\bibitem{Kress2007} Kress J.M., Equivalence of superintegrable systems in
  two dimensions, {\it Phys. Atomic Nuclei} {\bf 70} (2007), 560--566.

%-----------------------------------------------------------------------


\bibitem{AAR}
Andrews G.E., Askey R., Roy R.,
Special functions, {\it Encyclopedia of Mathematics and Its Applications},
Cambridge University Press,\ Cambridge, UK, 1999.
%-----------------------------------------------------------------------
\bibitem{KALNINS} Kalnins E.G.,
Separation of variables for Riemannian spaces of constant
curvature, {\it Pitman, Monographs and Surveys in Pure and Applied Mathematics}, Vol.~28,
 Longman, Essex, England,
 1986.
%-----------------------------------------------------------------------


\bibitem{ARNOLD}
Arnold V.I.,
Mathematical methods of classical mechanics
(translated by K.~Vogtmann and A.~Weinstein)
{\it Graduate Texts in Mathematics}, Vol.~60,
 Springer-Verlag,
 New York, 1978.
%-----------------------------------------------------------------------




\end{thebibliography}
\end{document}